\documentclass[3p,times]{elsarticle}

\usepackage{ecrc}

\volume{00}

\firstpage{1}

\journalname{Nuclear Physics A}

\runauth{}

\jid{npa}

\jnltitlelogo{Nuclear Physics A}

\usepackage{amssymb}
\usepackage{atlasphysics}

\usepackage{lineno}

\biboptions{square,comma,numbers,sort&compress}

\usepackage[figuresright]{rotating}

\newcommand{\FCalLowEta}{\mbox{$3.1$}}

\newcommand{\dndeta}{\mbox{$dN_{\mathrm{ch}}/d\eta$}}
\newcommand{\energy}{\mbox{$\sqrt{s_{NN}}=5.02$~\TeV}}
\newcommand{\sqn}{\mbox{$\sqrt{s_{NN}}$}}
\newcommand{\sqs}{\mbox{$\sqrt{s}$}}

\newcommand{\PbPb}{\mbox{Pb+Pb}}
\newcommand{\pPb}{\mbox{$p$+Pb}}

\newcommand{\pp}{\mbox{$p$+$p$}}

\newcommand{\sigNN}{\mbox{$\sigma_{\mathrm{NN}}$}}

\newcommand{\Npart}{\mbox{$N_{\mathrm{part}}$}}
\newcommand{\avgNpart}{\mbox{$\langle N_{\mathrm{part}}\rangle$}}

\newcommand{\sumETPb}{\mbox{$\Sigma E_{\mathrm{T}}^{\mathrm{Pb}}$\normalsize}}

\newcommand{\OCap}{\mbox{$\Omega$}}

\newcommand{\ystar}{\mbox{$y^{\mathrm{\star}}$}}

\newcommand{\rpPb}{\mbox{$R_{\mathrm{pPb}}$}}

\newcommand{\Gl}{Glau\-ber}
\newcommand{\GG}{Glau\-ber-Gri\-bov}

\newcommand{\Dp}{\mbox{$\Delta\phi$}}
\newcommand{\Yp}{\mbox{$Y(\Dp)$}}
\newcommand{\pta}{\mbox{$p_{\mathrm{T}}^{a}$}}
\newcommand{\ptb}{\mbox{$p_{\mathrm{T}}^{b}$}}

\begin{document}
\begin{frontmatter}


\dochead{}

\title{Particle production and long-range correlations in \pPb\ collisions with the ATLAS detector.}


\author{Alexander Milov on behalf of the ATLAS Collaboration}
\address{Department of Particle Physics and Astrophysics, Weizmann Institute of Science, 234 Herzl str., Rehovot 7610001, Israel}

\begin{abstract}
The ATLAS experiment at the LHC has measured the centrality dependence of charged particle pseudorapidity distribution, charged particle spectra, and the two-particle correlations in \pPb\ collisions at a nucleon-nucleon centre-of-mass energy of \sqn=5.02\,\TeV. Charged particles were measured over $|\eta|<2.7$ using the ATLAS detector tracking system. The \pPb\ collision centrality was characterized by the total transverse energy deposited over the interval $3.2<\eta<4.9$ in the direction of the Pb-beam.  Three different calculations of the number of participating nucleons, have been carried out using a standard \Gl\ initial state model as well as two \GG\ extensions. Charged particle multiplicities per participant pair, and the normalised charged particle spectra are found to vary in shape with $\eta$ and also with the model, pointing to the importance of the fluctuating nature of nucleon-nucleon collisions in the modelling of the initial state of \pPb\ collisions. The two particle correlation exhibits flow-like modulations for all centrality intervals and particle \pT.

\end{abstract}


\end{frontmatter}

\section{Introduction}
\label{sec:intro}

Proton-lead collisions (\pPb) at the Large Hadron Collider (LHC) provide an opportunity to probe the physics of the initial state of ultra-relativistic heavy ion collisions without obscuring effects of thermalization and collective evolution \cite{Csernai:2006zz}. The LHC delivered its first \pPb\ collisions at \sqn=5.02\,\TeV\ in a short pilot run in September 2012. Over several hour ATLAS detector~\cite{Aad:2008zzm} collected an event sample with integrated luminosity of approximately $1~\mu{\rm b}^{-1}$, assuming an input cross section for \pPb\ collisions of 2.1b estimated using a \pPb\ \Gl\ model.
  
This report reviews the results on the measurement of the charged particle production and correlations observed in \pPb\ collisions at the LHC. For the first time the results are analysed as a function of centrality, which provides additional information for understanding the underlying physics. Results are presented for several centrality intervals characterised by the sum of transverse energy (\sumETPb) measured in the forward (Pb-going) section of the ATLAS calorimeter.

\section{Centrality determination in \pPb\ collisions}
\label{sec:centr}

\begin{figure}[!htb]
\includegraphics[width=0.46\textwidth]{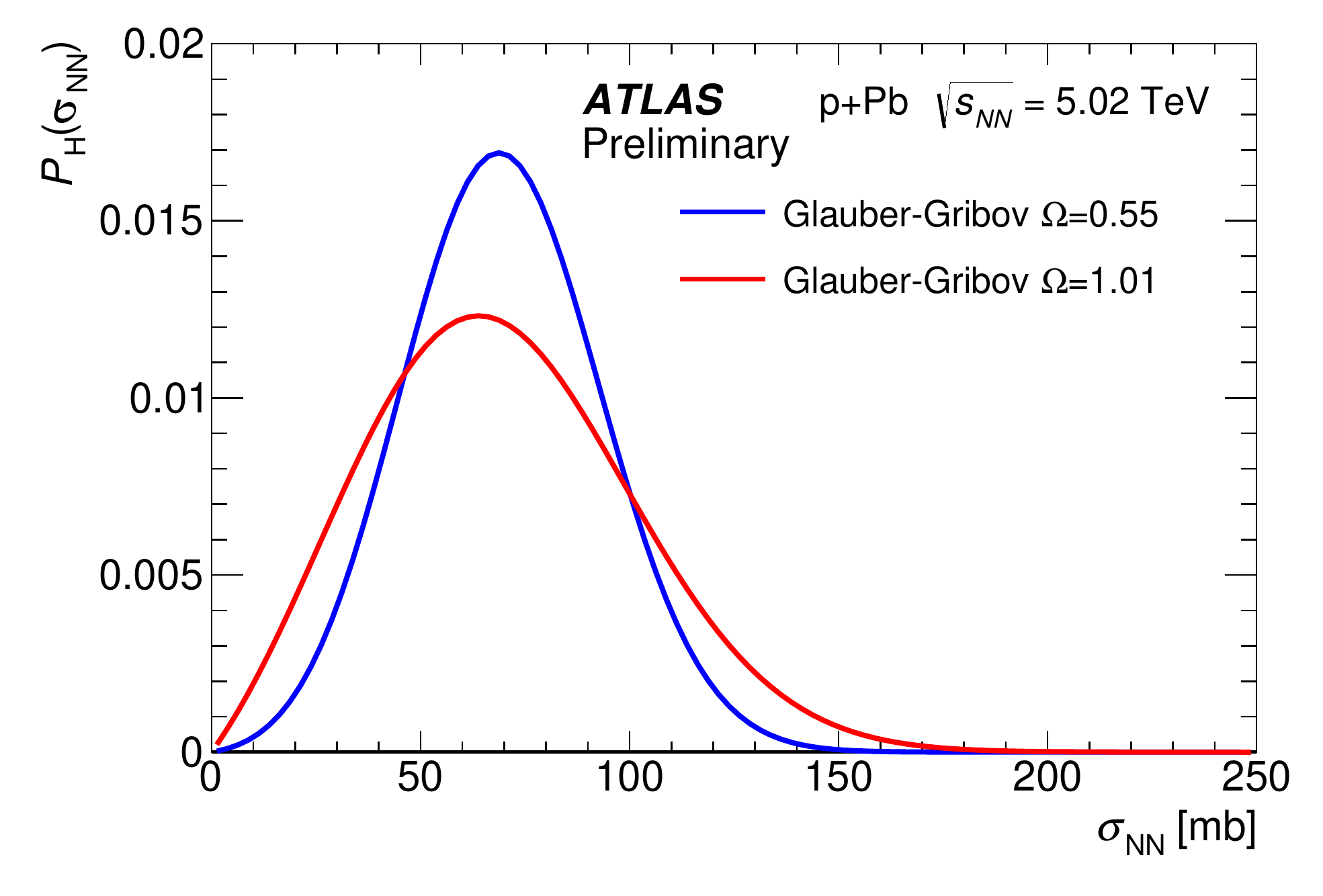} \vspace{-50mm} \\
\includegraphics[width=0.46\textwidth]{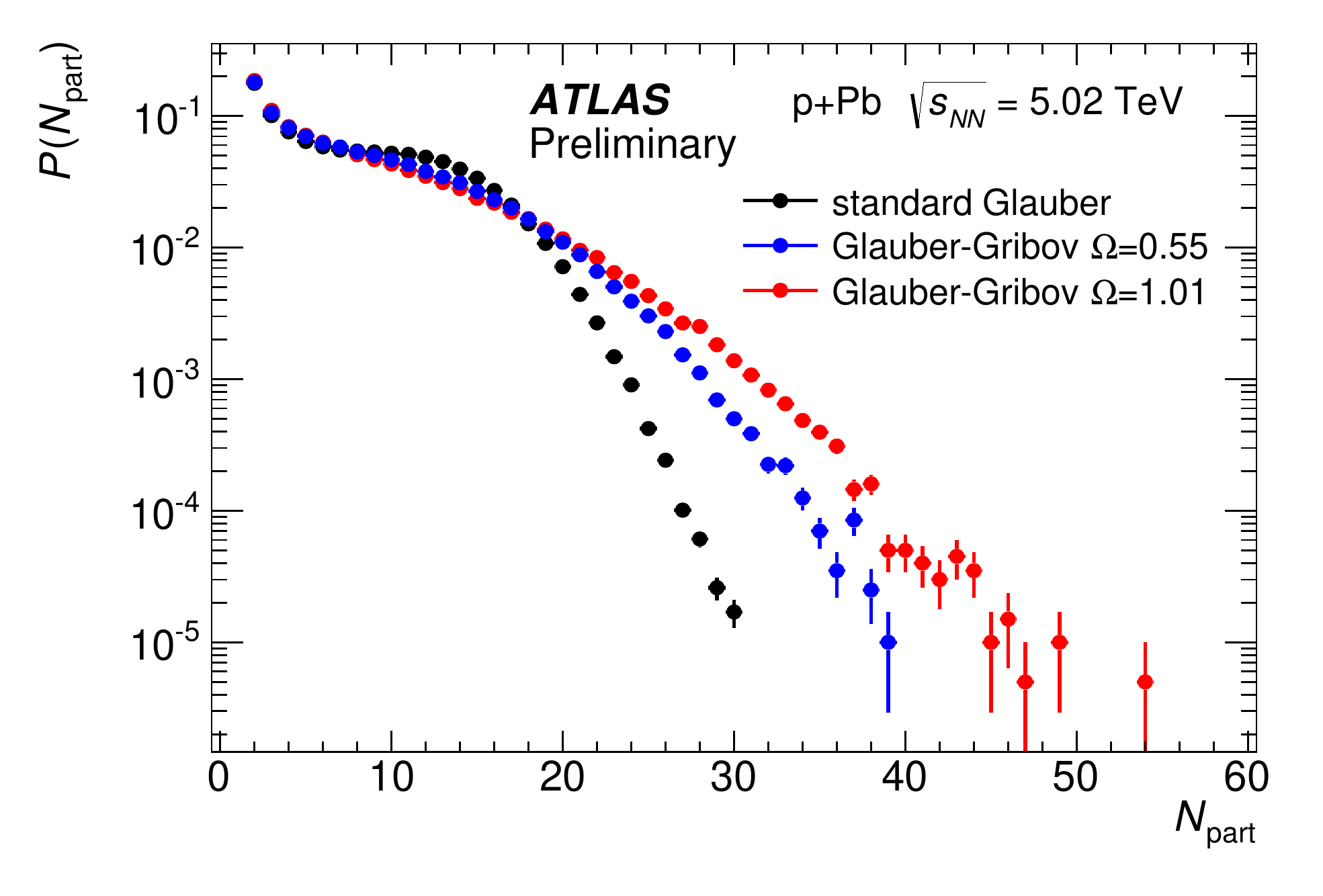} \vspace{-24mm}
\includegraphics[width=0.52\textwidth]{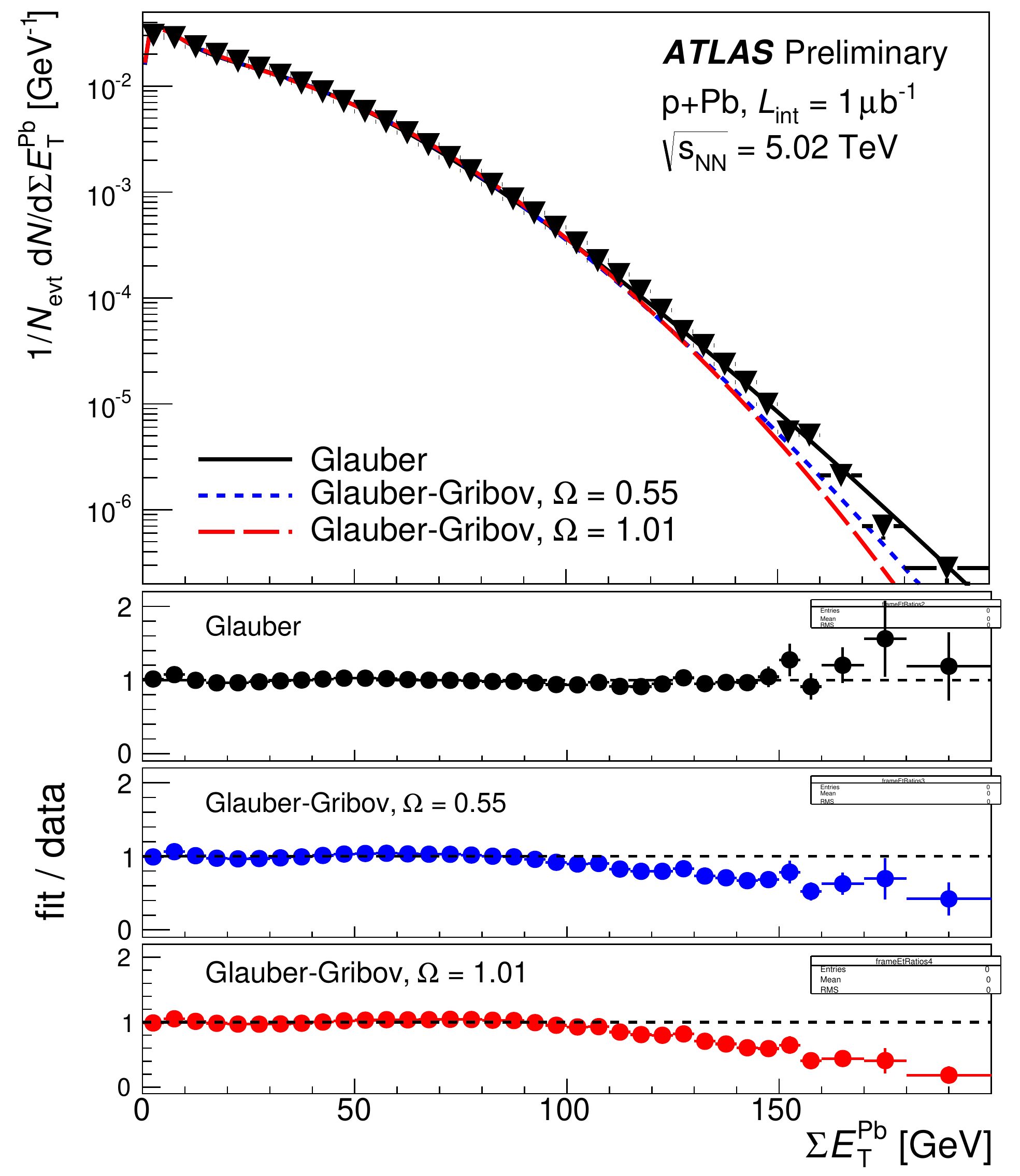}   \vspace{22mm}
\caption{Top left: \GG\ probability distribution for per-nucleon cross section for $\Omega = 0.55$ and 1.01. Lower left: The \Npart\ distributions for the \Gl\ and \GG\ Monte Carlo models simulated 5.02~\TeV\ \pPb\ collisions (1~million simulated events each). Top right: Comparison of the measured \sumETPb\ distribution to \Gl\ and \GG\ fits. Lower panels show ratios of the fits to the data distribution. Plots are from Ref.~\cite{ATLAS-CONF-2013-096}.}
\label{fig:GG}
\end{figure}

The geometry of nucleus-nucleus and proton-nucleus collisions can be studied using \Gl\ Monte Carlo model \cite{Alver:2008aq} that simulates interactions of incident nucleons using a semi-classical eikonal approximation. A modified version of the \Gl\ model, referred to as \GG, was implemented in the PHOBOS generator~\cite{Miller:2007ri} to account for event-to-event fluctuations in the nucleon-nucleon cross section (\sigNN)~\cite{Heiselberg:1991is,Guzey:2005tk,Alvioli:2013vk}. The \Gl\ model uses a fixed inelastic cross section, which is assumed to be 70~mb in this measurement. The \GG\ extentions use the probability distribution of the total cross section shown in the upper left panel of Fig.~\ref{fig:GG}. 
Parameter $\Omega$ defines the width of the total cross section probability distribution. The distribution of the number of nucleons participating in collision (\Npart) for the \Gl\ and \GG\ Monte Carlo models are shown in the lower left panel of the figure. The \GG\ analysis was performed using both $\OCap\ = 0.55$ and 1.01, motivated in Refs.~\cite{Guzey:2005tk,PhysRevLett.111.012001,Donnachie:1992ny,Alvioli:2013vk}, in order to evaluate the sensitivity of the physics conclusions to the choice of \OCap. For all three calculations, the lead nucleon density distribution was taken to be Woods-Saxon with radius and skin depth parameters, $R = 6.62$~fm and $a = 0.546$\,fm respectively~\cite{DeJager:1987qc}. The \GG\ \Npart\ distributions are much broader than the \Gl\ distribution due to \sigNN\ fluctuations in the \GG\ formulation.

The Wounded Nucleon (WN) model~\cite{Bialas:1976ed} is used to connect an experimentally measured \sumETPb\ distribution to the results of the \Gl\ or \GG\ Monte Carlo. For fixed \Npart, the measured distribution can be obtained from the WN model as an $n$-fold convolution of the corresponding \pp\ distribution, where $n$ is equal to \Npart.  The \pp\ distribution is taken from a full detector simulation based on PYTHIA8 (and verified with PYTHIA6)~\cite{mctunes} at \sqs=5.02\,\TeV. The results of the simulation were compared at the detector level to the measured distributions at lower \sqs=2.76\,\TeV\ and higher 7\,\TeV\ \pp\ energies. The distributions are approximated by the gamma distribution form ${\rm gamma}(k, \theta)$. The $n$-fold convolution of the gamma distribution is straightforward to implement analytically, however,  to obtain the best agreement between the data and the models the parameters of the distributions were assumed to be dependent on the number of participants: $k(\Npart)$, $\theta(\Npart)$. The results of the fit using the \Gl\ or \GG\ Monte Carlo to reproduce the measured \sumETPb\ distribution are shown in the right panel of Fig.~\ref{fig:GG}. The fit allows to associate the energy interval of \sumETPb\ to the \avgNpart\ from the \Gl\ model.

\section{Charged particle multiplicity}
\label{sec:mult}
\begin{figure}[!htb]
\includegraphics[width=0.40\textwidth]{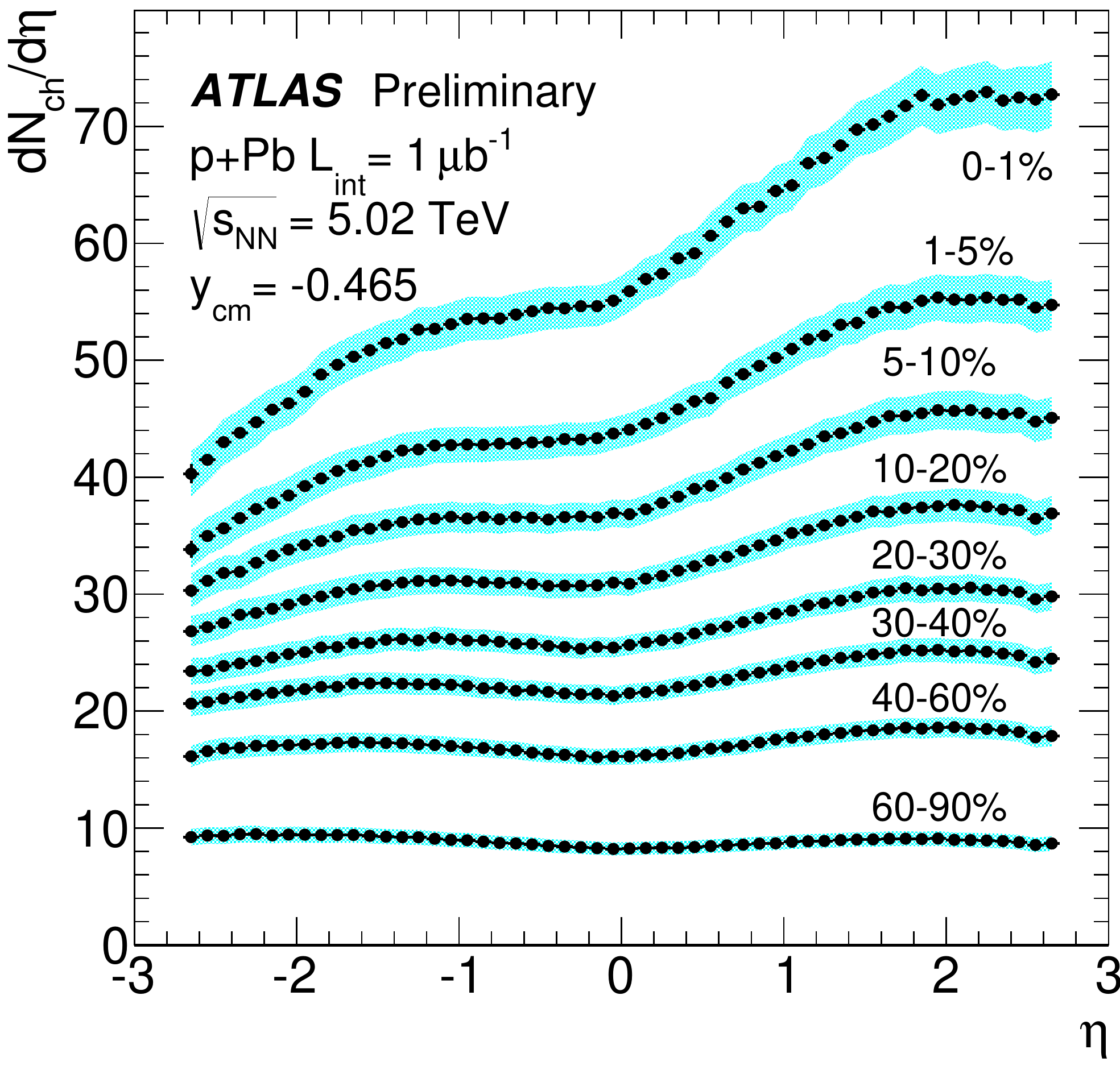} \vspace{-64mm} \\
\includegraphics[width=0.40\textwidth]{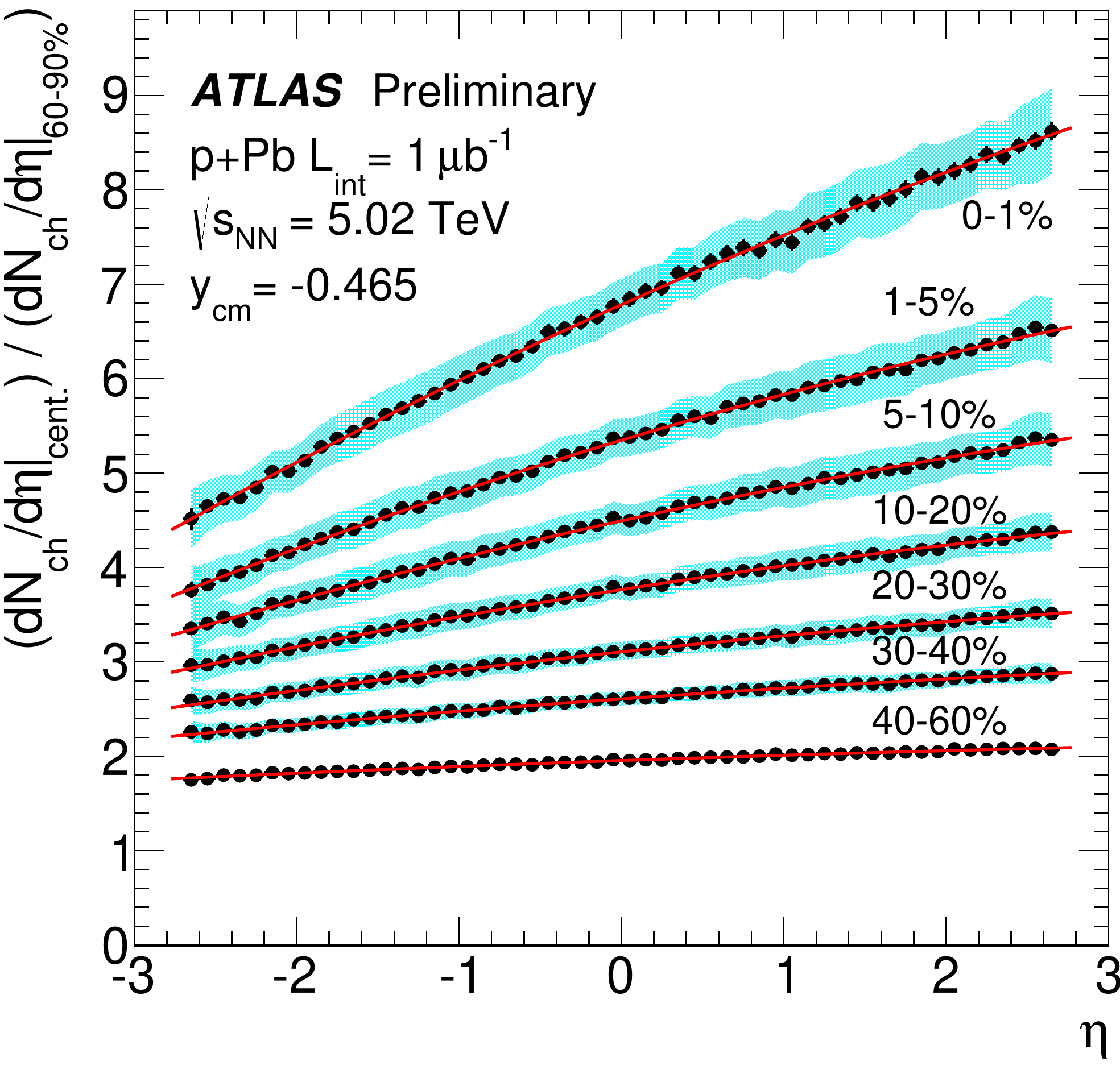} \vspace{-24mm}
\includegraphics[width=0.56\textwidth]{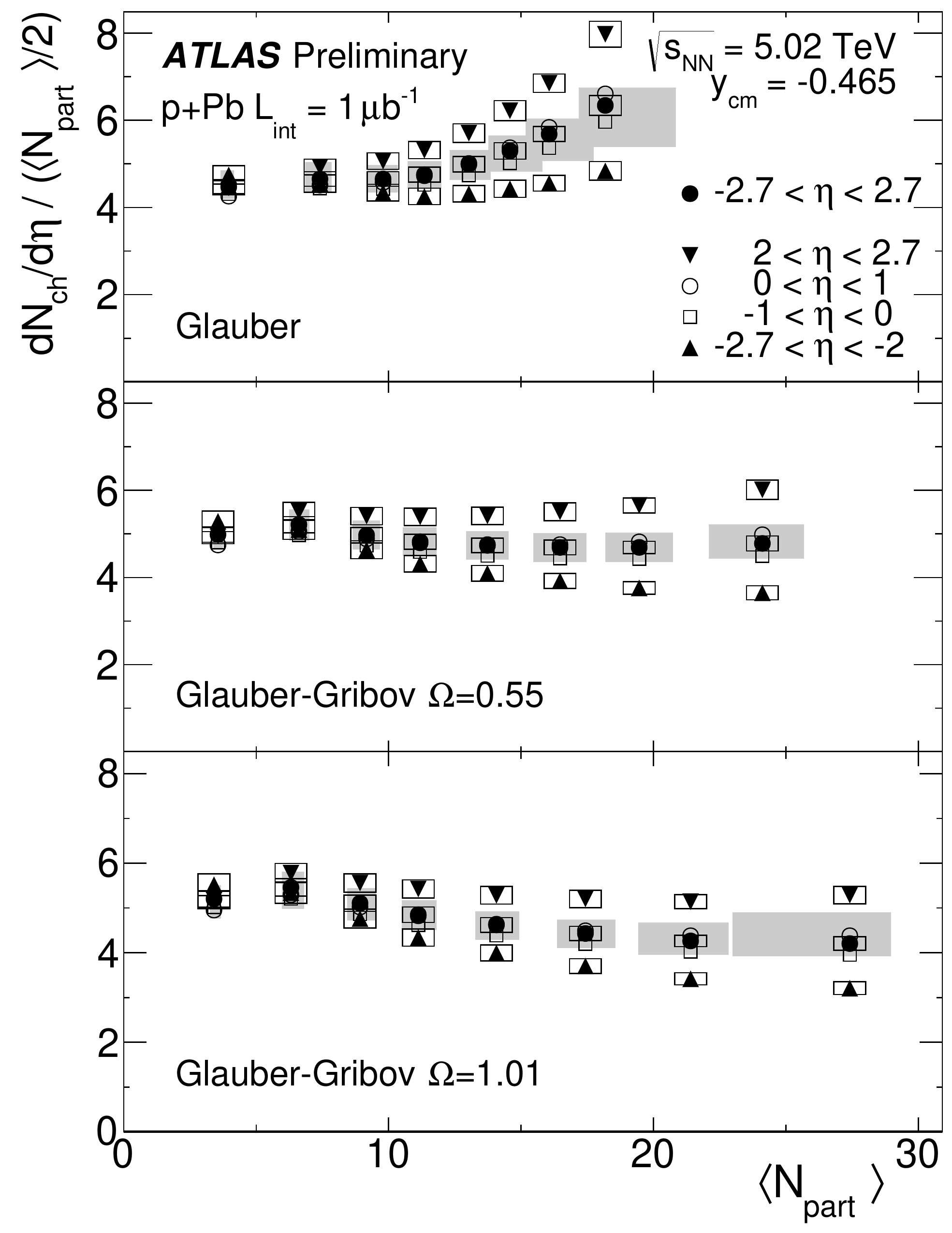}  \vspace{20mm}
\caption{Top left: Charged particle density \dndeta\ measured in different centrality intervals. Statistical uncertainties, shown with vertical bars, are typically smaller than the marker size. Colour bands show the systematic uncertainty. Bottom left: The \dndeta\ measured in different centrality classes divided by \dndeta\ measured in the peripheral (60-90\%) centrality interval. Lines show the results of second order polynomial fits to the data points. Right panels: Charged particle density \dndeta\ per pair of participants 
measured in different $\eta$ ranges, as a function of \avgNpart\ obtained using three initial state models.
The bands shown with thin lines represent the systematic uncertainty of the \dndeta\ measurement, the shaded bands indicate the total systematic uncertainty including the uncertainty on \avgNpart. Statistical uncertainties, shown with vertical bars are typically smaller than the marker size. Plots are from Ref.~\cite{ATLAS-CONF-2013-096}.}
\label{fig:mult}
\end{figure}

The measurement of the charged particle multiplicity was performed using only the pixel detector to maximize the efficiency for reconstructing charged particles with low-transverse momenta~\cite{ATLAS-CONF-2013-096}. Two approaches are used in this analysis. The first is the two-point tracklet method commonly used in heavy ion collision experiments~\cite{ATLAS:2011ag,Adcox:2000sp,Alver:2010ck}. The second method uses ``pixel tracks'', obtained by applying the full track reconstruction algorithm to only the pixel detector. The pixel tracking is less efficient than the tracklet method, but provides measurements of particle \pT.  The \dndeta\ measured using pixel tracks is used as a validation of the measurement with the two-point tracklets. The tracklet method used in two different implementations and the pixel track methods agree to each other. The first results on the \dndeta\ measured in different centrality intervals are shown in Fig.~\ref{fig:mult}.

The upper left panel of Figure~\ref{fig:mult} presents the charged particle pseudorapidity density for \pPb\ collisions at \energy\ over the pseudorapidity interval $|\eta|<2.7$ in eight centrality intervals from the 0--1\% most central to 60--90\% most peripheral. In the most peripheral collisions the \dndeta\ has a characteristic shape which is similar to a symmetric double-peak structure, seen in proton-proton collisions~\cite{Aad:2010ac,Aad:2010rd}. In more central collisions, the shape of \dndeta\ becomes progressively more asymmetric, with more particles produced in the Pb-going direction than in the proton-going direction. 

To investigate further the centrality evolution, the distributions in the various centrality intervals are divided by the distribution measured in 60-90\% centrality. The ratios are shown in the lower left panel of the figure. The double peak structure seen in the upper panel disappears in the ratios. The ratios are observed to grow nearly linearly with pseudorapidity, with a slope that increases from peripheral to central collisions. The ratios are well approximated with a second order polynomial fitted to data points.

The right panel of the figure shows \dndeta\ per participant pair for the most central and most peripheral intervals of centrality measured as a function of \eta\ for three different implementations of the \Gl\ model explained in Sec.~\ref{sec:centr}. The results for the most peripheral (60--90\%) centrality interval shown with circles, reside approximately at the same magnitude in all three panels. In the region $-1<\eta<0$ this magnitude is consistent with the $(\propto s^{0.10})$ approximation of \dndeta\ in inelastic \pp\ collisions, shown in Fig.~1 of Ref.~\cite{ALICE:2012xs}. The magnitude of \dndeta\ per participant pair strongly depends on pseudorapidity, consistent with the findings at RHIC~\cite{Back:2004mr}, and also on the model used to calculate \avgNpart. 

\section{Nuclear modification factor}
\label{sec:rpPb}

The ATLAS experiment at the LHC measures the differential invariant \pT\ yields for charged particles in \pPb\ collisions at \sqs=5.02~\TeV~\cite{ATLAS-CONF-2013-107}. The spectra are reconstructed over the available pseudorapidity interval $|\eta|<2.5$ in the same centrality intervals as the multiplicity measurement. The spectra are studied as a function of the centre-of-mass rapidity $\ystar=y+0.465$, calculated assuming the particle having the pion mass, and then corrected for the actual mass based on the Monte-Carlo simulations. The fully corrected charged particle production cross section as a function of \pT\ are also measured in \pp\ collisions at the centre of mass energies of \sqs=2.76\,\TeV\ and 7\,\TeV. The differential cross sections at the two energies are interpolated to obtain the differential cross section at \sqs=5.02\,\TeV\ used to derive the nuclear modification factor (\rpPb). The definition of the \rpPb\ can be found in~\cite{ATLAS-CONF-2013-107}.

\begin{figure}[!htb]
\begin{center}      
\includegraphics[width=0.75\textwidth]{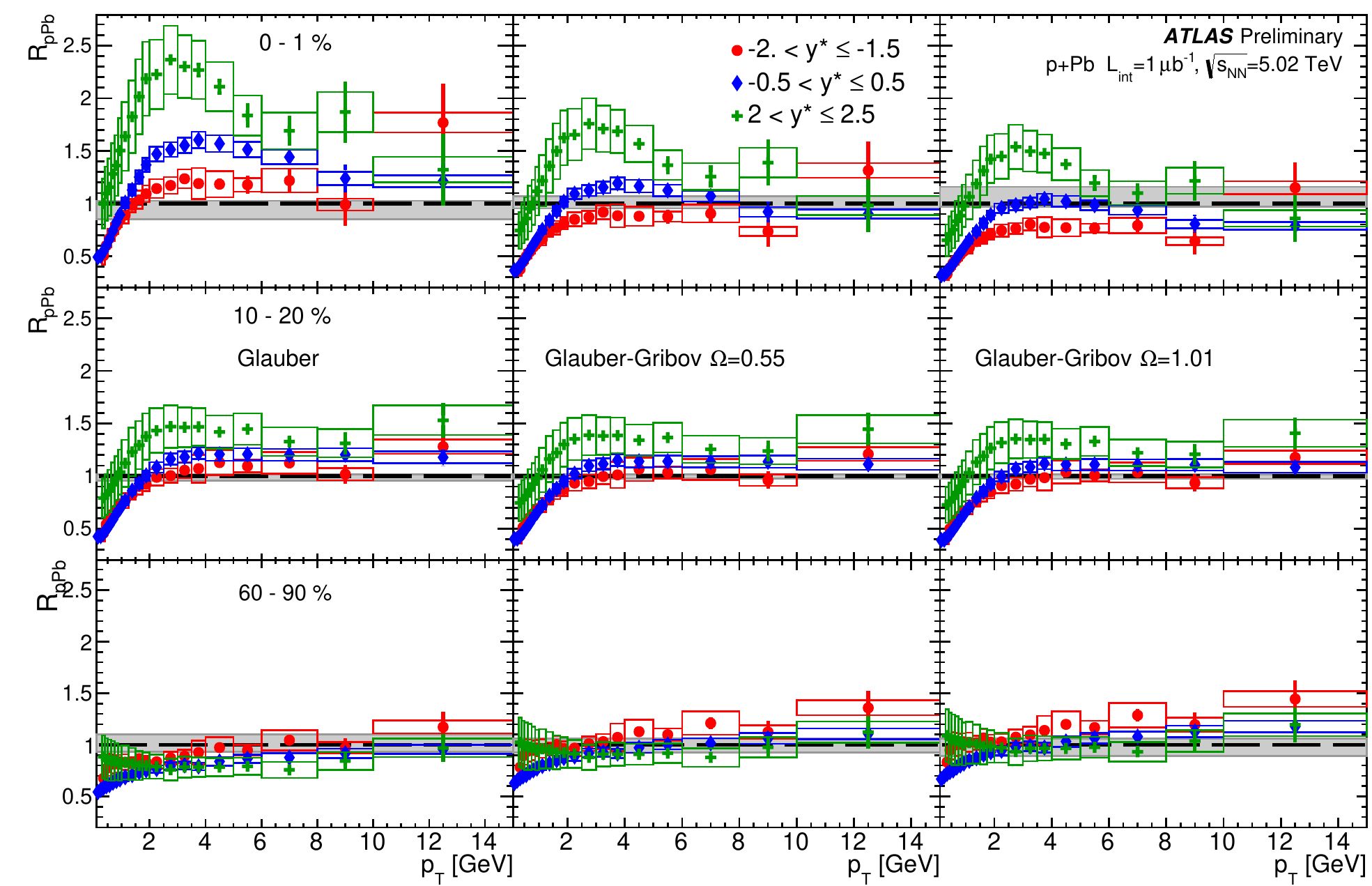}   
\caption{The \rpPb\ 
measured for 0-1\% most central collisions (top panels) 10-20\% centrality interval (middle panels) and 60-90\% centrality interval (lower panels). The data points are shown for 3 different rapidity intervals indicated in the legend. The columns correspond to the \Gl\ model (left), \GG\ model with $\Omega=0.55$ (middle), and \GG\ model with $\Omega=1.01$ (right) calculations. The grey box on each axis reflects the systematic uncertainty associated with the centrality interval and with the model assumption. Systematic uncertainties due to the analysis of the spectra are plotted with boxes at each point. Vertical bars are the statistical uncertainty of the measurement. Plot is from Ref.~\cite{ATLAS-CONF-2013-107}.} 
\label{fig:rcp_EtaBins_CBins}
\end{center}
\end{figure}

Figure~\ref{fig:rcp_EtaBins_CBins} shows \rpPb 
as a function of \pT. Three panels in each column correspond to the most central (upper panels), mid-central (middle panels) and most peripheral (lower panels) centrality intervals. Each panel has three sets of points corresponding to different \ystar\ intervals for \rpPb. The three columns show the results from the three different geometrical models: \Gl\ model (left), \GG\ model with $\Omega=0.55$ (middle), and \GG\ model with $\Omega=1.01$ (right). 

Figure~\ref{fig:rcp_EtaBins_CBins} shows that  the magnitude of the peak at \pT=2-4\,\GeV, relative to the almost constant behaviour above 8\,\GeV, depends 
not only on centrality, but also on rapidity. It increases with increasing \ystar\ (i.e. moving farther toward the rapidity of the Pb nucleus), as observed in the upper and middle panels of the figure. The rapidity intervals in the proton beam direction (negative \ystar) show a smaller peak compared to the rapidity interval in the Pb beam direction. The magnitude of the constant region also depends on rapidity. The nuclear modification factors in the constant region ($\pT > 8$ GeV) show much less variation over the measured rapidity region for all centralities. The limited statistics of the pilot \pPb\ run precludes conclusive statements about the \rpPb\ behaviour as a function of \ystar\ in the high \pT\ part of the spectrum.

\section{Charged particle correlations}
\label{sec:corr}

The ATLAS experiment measured the strength of the long-range two-particle correlations in \pPb\ collisions at \sqs=5.02\,\TeV\ quantified by the ÔÔper-trigger yieldÕÕ, (\Yp)~\cite{Aad:2012gla}. The yields are integrated over relative differences $2<|\Delta\eta|<5$ between the "trigger particle" ($a$) and associated particle ($b$) and the background is subtracted based on the zero-yield-at-minimum (ZYAM) method. 

\begin{figure}[!htb]
\begin{center}
\includegraphics[width=0.52\textwidth]{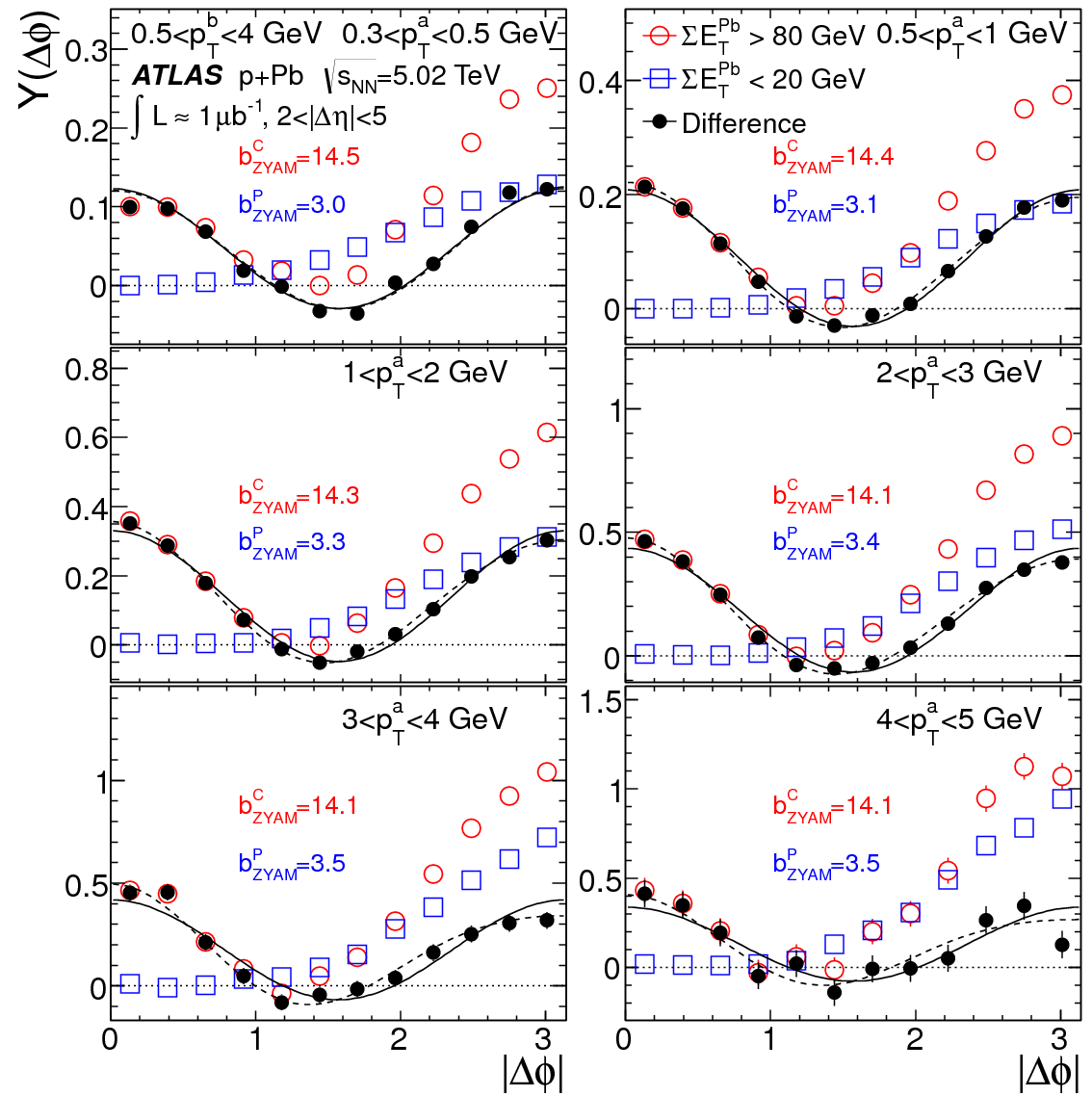}    
\includegraphics[width=0.45\textwidth]{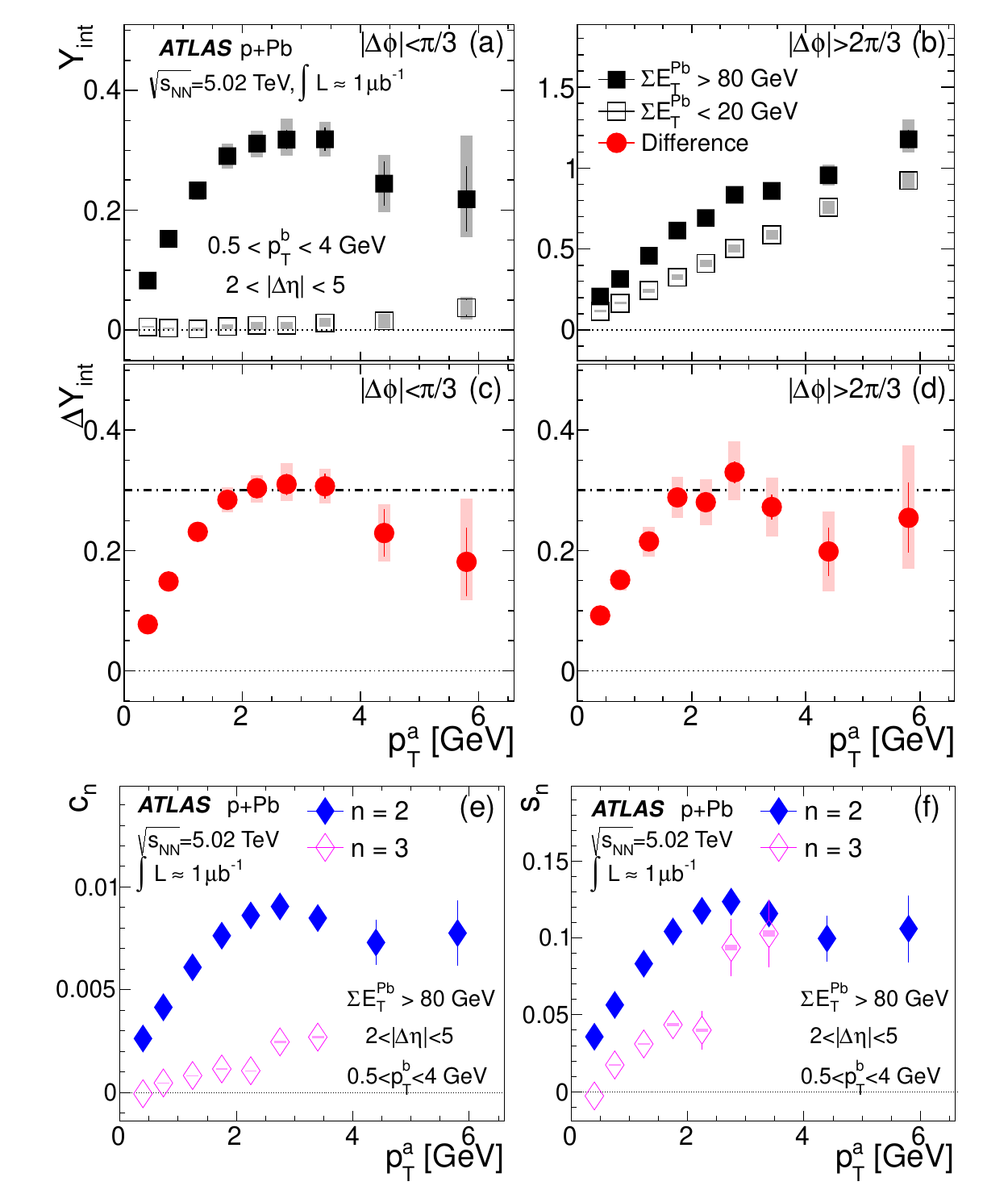}       
\caption{Left panels: Distributions of per-trigger yield in the peripheral and the central event activity classes and their differences (solid symbols), for different ranges of $\pT^{a}$ and $0.5<\pT^{b}<4$\,\GeV, together with functions $a_{0}+2a_{2}cos2\Delta\phi$ (solid line) and $a_{0}+2a_{2}cos2\Delta\phi+2a_{3}cos3\Delta\phi$ (dashed line) obtained via a Fourier decomposition. The values for the ZYAM-determined pedestal levels are indicated on each panel for peripheral ($b^{P}_{\mathrm{ZYAM}}$) and central ($b^{P}_{\mathrm{ZYAM}}$) \sumETPb\ bins.
Right panels: Integrated per-trigger yields, $Y_{\mathrm{int}}$, vs \pta\ for $0.5<\ptb<4$\,\GeV\ in peripheral and central events, on the (a) near-side and (b) away-side. The panels (c) and (d) show the difference, $\Delta Y_{\mathrm{int}}$. Panels (e) and (f) show the \pt\ dependence of $c_2$, $c_3$ and $s_2$, $s_3$. The error bars and shaded boxes represent the statistical and systematic uncertainties, respectively. Plots are from Ref.~\cite{Aad:2012gla}. 
} 
\label{fig:corr}
\end{center}
\end{figure}

\Yp\ distributions for peripheral and central events and their difference are shown as a function of $|\Delta\phi|$ in the left two columns of Fig.~\ref{fig:corr} 
for various \pta\ ranges with $0.5<\ptb<4$\,\GeV. The difference is observed to be nearly symmetric around $\Dp=\pi/2$. To illustrate this symmetry, the \Yp\ distributions in Fig.~\ref{fig:corr} are overlaid with the function symmetric around $\pi/2$ (solid curve), indicating that the long-range component of the two-particle correlations can be approximately described by a recoil contribution plus a \Dp-symmetric component. 

The near-side and away-side yields, integrated respectively over $|\Dp|\leq \pi/3$ and $|\Dp|>2\pi/3$ ($Y_{\mathrm{int}}$), and the differences between these integrated yields in central and peripheral events ($\Delta Y_{\mathrm{int}}$), are shown in the right two columns of Fig.~\ref{fig:corr} as a function of \pta. The yields are shown separately for two \sumETPb\ ranges in panels (a) and (b) and the differences are shown in panels (c) and (d). Qualitatively, the differences have a similar \pta\ dependence and magnitude on the near-side and away-side; they rise with \pta\ and reach a maximum around 3-4\,\GeV. 

The relative amplitude of the $cos(n\Dp)$ modulation of \Yp, ($c_n$) can be estimated from the fits shown in the left two columns as explained in Ref.~\cite{Aad:2012gla}. Panel (e) of Fig.~\ref{fig:corr} shows $c_2$ and $c_3$ as a function of \pta\ for $0.5<\ptb<4$\,\GeV. The value of $c_2$ is much larger than $c_3$ and exhibits a behaviour similar to \Yp\ at the near-side and away-side. The values of $c_n$ can be converted into an estimate of $s_n$, the average $n$-th Fourier coefficient of the event-by-event single-particle $\phi$-distribution. The $s_2(\pta)$ values obtained this way exceed 0.1 at $\sim$2-4\,\GeV, as shown in Fig.~\ref{fig:corr}(f). The transverse momentum dependence of $s_2$ and $s_3$ is similar to that observed for long-range correlations in \PbPb\ collisions~\cite{ATLAS:2012at,10.1140/epjc/s10052-012-2012-3}.

\section{Summary}
\label{sec:sum}

This report presents the measurements of the centrality dependence of the charged particle pseudorapidity distribution,  \dndeta, \rpPb\ and two particle correlations in \pPb\ collisions as a function of \pT\ at a nucleon-nucleon centre-of-mass energy of \energy\ collected by the ATLAS detector using data sample of the pilot run commenced by the LHC in September 2012. The centrality is characterized using a forward calorimeter covering $\FCalLowEta < \eta < 4.9$ in the Pb-going direction.  The average number of participants in each centrality interval, \avgNpart, was estimated with the \Gl\ and \GG\ Monte Carlo models using the simulated response of the ATLAS forward colorimeter in Pb-going direction.

The \dndeta\ distribution, presented over the pseudorapidity range $-2.7<\eta<2.7$ evolve gradually with centrality from an approximately symmetric shape in the most peripheral collisions to a highly asymmetric distribution in the most central collisions. The ratios of $\dndeta$ distributions in different centrality intervals to the \dndeta\ in the most peripheral interval are approximately linear in $\eta$ with a slope that is strongly dependent on centrality.  The nuclear modification factor, \rpPb, is measured as a function of the centre-of-mass rapidity in the range $-2<\ystar<2.5$ and for transverse momentum $0.1<\pT<22$\,\GeV. The reference \pp\ cross-section is obtained by interpolating the measurements performed at \sqs=2.76\,\TeV\ and 7\,\TeV. The shape and the magnitude of the \rpPb\ was found to have a strong rapidity dependence. Both the \Npart\ dependence of $\dndeta/(\avgNpart/2)$ and the \rpPb\ are sensitive to the \Gl\ modeling, especially in the most central collisions. 

The two-particle correlation function measured in \pPb\ collisions clearly shows ridge structures resembling those observed in \PbPb\ collisions and suggesting that collective flow may also be present in \pPb\ collisions. The flow interpretation of the \pPb\ data is also supported by results from multi-particle azimuthal correlation measurements~\cite{Aad:2013fja}. 

The work of the author is supported by the Israel Science Foundation (grant 710743).

\bibliographystyle{elsarticle-num}
\bibliography{sasha_milov_proceeding_copy.bbl}







\end{document}